# Теоретический предел сжатия информации

*Copyright 2000 © Лаврёнов А.*

*E-mail: lanin99@mail.ru*

Abstract: The pit recording of file, the coefficient of compression are introduced. The theoretical limit of the information compression as minimal coefficient of compression for the given length of alphabet are found.

Термин «сжатие информации» может употребляться в разных смыслах. Использование его в данной статье лучше понять при помощи философских категорий содержание и форма. Упорядочивание и классификация информации на первых порах, затем создание различных моделей и теорий можно рассматривать как смысловое или содержательное сжатие информации. В частности, этот процесс особенно ярко проявляется в естественных науках. Так, считается, что уравнения Максвелла являются теоретической основой существующей техники. Однако такой вид сжатия не решает в принципе проблему сжатия полученной информации. В последнее время становится важным сохранение уникальных документов, количество которых растет по экспоненциальному закону. В частности, книги или картины могут воспроизводиться на различных носителях и, соответственно, с разным занимаемым физическим объемом. Данный вид сжатия определим как сжатие информации по форме, что и будет очерчивать наш круг интересов в философском плане. Отметим, технологические вопросы увеличение плотности информации на ее носителях и миниатюризации не будут обсуждаться в данной статье.

Вначале дадим несколько определений.

**Алфавит (А)** – конечный набор символов любой природы.

**Текст** – любая последовательность символов, возможно с повторениями.

**Основной текст** – последовательность всех символов алфавита без повторения.

**Пит** – величина, принимающая ровно *p*-значений.

**N-разрядный кортеж** – питовая последовательность, состоящая из N пит.

**Алфавит разрядности N ($A_n$)** – множество всех различных N-разрядных кортежей.

**Файл** - любая последовательность N-разрядных кортежей, возможно с повторениями.

**Основной файл** - последовательность всего алфавита разрядности N ($A_n$) без повторения.

**Длина файла** – количество пит, используемых для представления файла.

**Питовая перекодировка алфавита разрядности N ($A_n$)** – взаимооднозначное отображение между множеством алфавита разрядности N ($A_n$) с постоянной длиной кортежа *n* и конечным множеством алфавитов $A_k$ с длиной кортежа $k \le n$.

**Коэффициент сжатия** $k_{\min} = \dfrac{L_2}{L_1}$ - отношение длин основного файла после $L_2$ и до $L_1$ питовой перекодировки алфавита разрядности N ($A_n$)

**Цифровое представление символа, алфавита (A), текста** – это соответственно N-разрядный кортеж, алфавит разрядности N ($A_n$), файл при условии взаимооднозначного соответствия алфавита (A) и алфавита разрядности N ($A_n$).

Известно, что для представления любой информации можно выбрать множество обьектов-кирпичиков, из которых строится все. В частности, а) с помощью трех цветов (красный, зеленый, синий) можно получить все цвета радуги; б) любое литературное произведение записывается только при помощи алфавита соответствующего языка и т.д. Поэтому данное множество обьектов-кирпичиков назовем алфавитом, а такое представление - алфавитным. Использование вычислительной техники диктует цифровое представление алфавита, которое основывается в настоящее время на позиционной системе исчисления. Но оно избыточно по двум причинам. Во-первых, общее число символов алфавита (A) предполагается равным $p^n$, где ***p*** - это число-основание системы исчисления, ***n*** – число-разрядность, или длина кортежа. Во-вторых, любой символ алфавита (A) имеет постоянную длину кортежа n. Также специально отметим преобладающее использование двоичной системы исчисления в существующей вычислительной технике. Чтобы не привязываться к конкретному виду информации или к конкретному документу, далее будем рассматривать только основной текст или основной файл. Цель настоящей статьи связана с нахождением минимального коэффициента сжатия $k_{min} = \dfrac{L_2}{L_1}$ при данной длине первоначального алфавита (А) $l_A$. Очевидно, что данная мера избыточности цифрового представления информации в разных системах исчисления покажет нам теоретический предел сжатия информации.

Пусть для представления всего алфавита А нам необходимо $p^n$ элементов, где **p** – это число-основание представления, **n** – число-разрядность или длина кортежа. Следовательно, длина $L_1$, занимаемая всеми элементами алфавита, в стандартном виде с постоянной длиной кортежа n равна

$$L_1 = np^n.$$

Для решения поставленной задачи нам необходимо установить взаимооднозначное соответствие между множеством алфавита А=$A_n$ с постоянной длиной кортежа n и конечным множеством алфавитов $A_k$ с длиной кортежа $k \leq n$. При p>1 длина $L_2$, занимаемая всем элементами алфавита, в сжатом виде или при наиболее полном использовании множества алфавитов $A_k$ с длиной кортежа $k \leq n$ имеет следующий вид:

$$\begin{aligned}L_2 = 1p^1 + 2p^2 + ..... + kp^k + .... + (n-1)p^{n-1} + \\ + n\left[p^n - \left(p^1 + p^2 + ... + p^{n-1}\right)\right]\end{aligned},$$

где в первой строке выписана длина элементов алфавита А=$A_n$ при замене их элементами алфавитов $A_k$, а во второй строке – оставшиеся элементы алфавита А=$A_n$.

Теоретический предел сжатия информации мы связываем с нахождением минимального коэффициента

$$k_{min} = \dfrac{L_2}{L_1}.$$

В данном случае он равен

$$k_{\min} = \frac{np^{n+2} - p^{n+1}(2n+1) + p^n n + p^2 n + p(1-n)}{np^n(p-1)^2}$$

Усложним условие задачи. Для представления всего алфавита (А) нам необходимо $l_A$ символов. В системе исчисления с основанием **p** число $l_A$ представляется в следующем виде $l_A = p^{n-1} + d$, где $0 < d \le p^n - p^{n-1}$. Следовательно, длина основного файла $L_1$ в стандартном виде с постоянной длиной кортежа n равна

$$L_1 = n[p^{n-1} + d].$$

При p>1 длина основного файла $L_2$ в сжатом виде или при наиболее полном использовании множества алфавитов $A_k$ с длиной кортежа $k \le n$ имеет следующий вид:

а) при $d \le S_{n-1} = \dfrac{p^{n-1} - p}{p - 1}$

$$L_2 = \frac{(n-2)p^n - (n-1)p^{n-1} + p}{(p-1)^2} + (n-1)[d + \frac{p^n - 2p^{n-1} + p}{p-1}];$$

б) при $d \ge S_{n-1}$

$$L_2 = \frac{(n-1)p^{n+1} - np^n + p}{(p-1)^2} + n[d - S_{n-1}];$$

Рассмотрим как оценки коэффициента сжатия согласуются с представлением информации в существующих вычислительных средствах. Обычно размер алфавита в них определяется числом 256, т.е. $l_A$=256.

$$256 \equiv 2^8 = 3^5 + d_3 = 4^4 = 5^3 + d_5 = 6^3 + d_6 =$$
$$= 7^2 + d_7 = ..... = k + d_k = ........ = 16^2 =$$
$$= 17^1 + d_{17} = ........ = 254^1 + 2 = 255^1 + 1 = 256^1$$

Соответственно, для каждого его представления будем иметь свои значения коэффициента сжатия. Ниже приведен только несколько интересных значений:

$k_{\min} \approx 0.76$ (p=2); 0.72 (p=3); 0.89 (p=4); 0.7 (p=6);
0.74 (p=13); 0.67 (p=15); 0.97 (p=16) и 1 (p=256);

Таким образом, наибольшее сжатие достигается не только при использовании двоичной системы исчисления. В заключение отметим, что вышеуказанную процедуру сжатия можно выполнить несколько раз (вплоть до получения длины, меньшей n). Однако в этом случае нам необходимо сохранять дополнительную информацию для однозначного восстановления информации. Результаты статьи докладывались на конференции "Наука и педагогика на рубеже 21 столетия", посвященной 10-летию образования ИСЗ (Беларусь, Минск, 26-27 октября 2000 г.)